\documentclass[aps,floatfix,prl,twocolumn,a4paper,10pt,showpacs,citeautoscript,
preprintnumbers,superscriptaddress]{revtex4-1} 

\usepackage[english]{babel}	

\usepackage{amsmath}		
\usepackage{amssymb}		
\usepackage{bbm}		

\usepackage{graphicx}		
\usepackage{graphics}		
\usepackage{pict2e}		
\usepackage{xcolor}		

\DeclareGraphicsExtensions{.png,.jpg,.pdf}	

\usepackage[colorlinks=true, pdfstartview=FitV,linkcolor=blue, citecolor=blue, urlcolor=blue]{hyperref}
\renewcommand{\theequation}{\arabic{equation}}

\makeatletter
\newcommand*{\Equation}{\@ifstar\sEquation\oEquation}
\newcommand{\sEquation}[1]{\begin{equation*}#1\end{equation*}}
\newcommand{\oEquation}[2]{  \begin{equation}\label{#1}#2\end{equation} }
\makeatother
\newcommand{\Align}[2]{\begin{align}\label{#1}#2\end{align}}

\newcommand{\bs}{\boldsymbol}

\newcommand{\Figref}[1]{Fig.~\ref{#1}}
\newcommand{\Eqref}[1]{\eqref{#1}}

\newcommand{\eg}{{e.g.~}}
 
\newcommand{\ie}{{i.e.~}}

\newcommand{\groupU}[1]{\mathrm{U}(#1)} 
\newcommand{\groupZ}[1]{\mathbb{Z}_{#1}} 
\newcommand{\groupCP}[1]{{\mathbbm{C}{P}}^{#1}} 
\newcommand{\CPtwo}{$\ \groupCP{2}\ $}
\newcommand{\Uone}{$\ \groupU{1}\ $}
\newcommand{\Ztwo}{$\ \groupZ{2}\ $}
\newcommand{\groupUZ}{$\ \groupU{1}\times\groupZ{2}\ $} 

\newcommand{\Grad}{\bs \nabla}

\newcommand{\Curl}{\bs \nabla\times}

\newcommand{\ez}{\bs e_z}

\newcommand{\F}{\mathcal{F}}
\newcommand{\G}{\mathcal{G}}

\newcommand{\SRO}{{Sr$_2$RuO$_4$\ }}

\newcommand{\Tc}{T_c}
\newcommand{\Tz}{T_{\groupZ{2}}  }
\newcommand{\Hc}[1]{\mathrm{H}_{c#1}} 
\newcommand{\HcO}{$\ \Hc{1}\ $}

\newcommand{\dT}{\delta T}
\newcommand{\dH}{\delta H}

\newcommand{\oz}{{(0)}}

\newcommand{\D}{{\bs D}}
\newcommand{\A}{{\bs A}}
\newcommand{\B}{{\bs B}}

\newcommand{\alphaZ}{\alpha^\oz}

\newcommand{\psia}{\psi_a}
\newcommand{\psib}{\psi_b}
\newcommand{\varphiab}{\varphi_{ab}}
\newcommand{\etaab}{\eta_{ab}}

\begin{document}
\title{Domain walls and their experimental signatures in 
\texorpdfstring{$s+is$}{s+is} superconductors}

\author{Julien~Garaud}
\email{garaud.phys@gmail.com}
\affiliation{Department of Physics, University of Massachusetts Amherst, MA 01003 USA }
\affiliation{Department of Theoretical Physics, Royal Institute of Technology, 
Stockholm, SE-10691 Sweden}
\author{Egor~Babaev}
\affiliation{Department of Physics, University of Massachusetts Amherst, MA 01003 USA }
\affiliation{Department of Theoretical Physics, Royal Institute of Technology, 
Stockholm, SE-10691 Sweden}
\date{\today}

\begin{abstract}

Arguments were recently advanced that  hole-doped 
Ba$_{1-x}$K$_x$Fe$_2$As$_2$  exhibits $s+is$ state at certain 
doping. Spontaneous breaking of time reversal symmetry in $s+is$ 
state, dictates that it possess domain wall excitations. Here, we 
discuss what are the experimentally detectable signatures of domain 
walls in $s+is$ state. 
We find that in this state the domain walls can have dipole-like
magnetic signature (in contrast to the uniform magnetic signature of 
domain walls $p+ip$ superconductors). We propose experiments where 
quench-induced domain walls can be stabilized  by geometric barriers 
and be observed via their magnetic signature or their influence on 
the magnetization process, thereby providing an experimental tool 
to confirm $s+is$ state.

\end{abstract}

\maketitle


The recently discovered iron-based superconductors 
\cite{Kamihara.Watanabe.ea:08} may exhibit new
physics originating in the possible frustration of inter-band 
couplings between more than two superconducting components 
\cite{Ng.Nagaosa:09,Stanev.Tesanovic:10,Carlstrom.Garaud.ea:11a,
Maiti.Chubukov:13}. 
For a two-band superconductor, inter-band Josephson interaction 
either locks or anti-locks phases, so that the ground state 
inter-band phase difference is respectively $0$ or $\pi$. 
Similarly, for more than two bands, each inter-band coupling 
favours (anti-)locking of the two corresponding phases. However, 
these Josephson terms can collectively compete so that optimal 
phases are neither locked nor anti-locked. There, the resulting 
\emph{frustrated} phase differences are neither $0$ nor $\pi$. 
Since it is not invariant under complex conjugation, such a ground 
state spontaneously breaks the Time-Reversal Symmetry (TRS)
\cite{Ng.Nagaosa:09,Stanev.Tesanovic:10}. 
This is the $s+is$ state, with the spontaneously Broken Time-Reversal 
Symmetry (BTRS), that recently received strong theoretical support 
in connection with hole-doped Ba$_{1-x}$K$_x$Fe$_2$As$_2$, 
\cite{Maiti.Chubukov:13}.
There are also other scenarios for BTRS states in pnictides 
\cite{Lee.Zhang.ea:09,Platt.Thomale.ea:12}, and related multi-component 
states may possibly exist in other classes of materials 
\cite{Mukherjee.Agterberg:11}. 

Symmetrywise, these BTRS states break the \groupUZ symmetry. The 
topological defects associated with the breakdown of a discrete 
\Ztwo symmetry are domain walls (DW) segregating regions of different 
broken states \cite{Manton.Sutcliffe}. Other superconductors with 
BTRS and having domain walls are the chiral $p$-wave superconductors.
There are evidences for such superconductivity in \SRO \cite{xia}.
For that material, it is predicted that domain walls have magnetic 
signature and thus can be detected by measuring the magnetic field 
(see \eg \cite{sigrist,goldbart}).
These signatures were searched for in surface probes measurements, 
but were not experimentally detected \cite{moler2}. This led to 
intense theoretical investigation of possible mechanisms for 
the field suppression (see \eg \cite{raghu}). The problem of 
interaction of vortices and domain walls in these systems and 
magnetization process was studied in \cite{machida1,machida2}.
Domain walls between BTRS states is also highly important in
rotational response of ${}^3He$ \cite{golov}.
Aspects of topological defects of $s+is$ states received 
attention only recently \cite{Garaud.Carlstrom.ea:11,
[{}][{. 
This work studied dynamics of unstable closed domain
walls and reported appearance of magnetic signatures.
We did not observe this kind of magnetic signatures in 
our simulations, which by contrast focuses on
(quasi-)equilibrium configurations.
} ]Lin.Hu:12a,
Garaud.Carlstrom.ea:13,Bojesen.Babaev.ea:13}. 
The remaining question is how domain walls can be created and 
observed in $s+is$ superconductors. In this paper, we demonstrate 
that these objects can be stabilized by geometric barriers in 
mesoscopic samples and discuss what experimental signatures it 
will yield.

\begin{figure}[!htb]
\vspace{-1.6cm}
\hbox to \linewidth{ \hss
\rotatebox{0}{\resizebox{.5\linewidth}{!}{\input{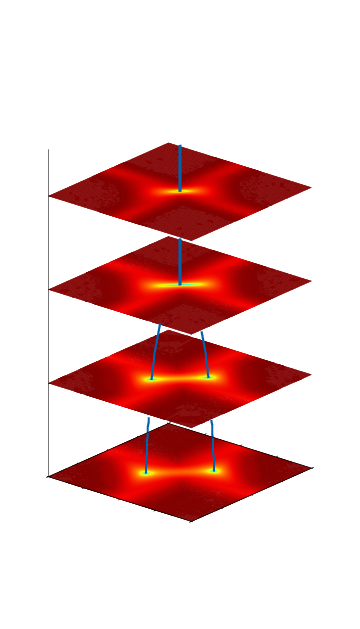}}}
\hss}
\vspace{-0.4cm}
\caption{
(Color online) -- 
This shows the symmetry breaking pattern for a frustrated 
three-band superconductor. Surfaces show the potential energy 
as a function of the phase differences, at different 
temperatures. The blue line shows the ground state. Above $\Tz$, 
phases are locked and the ground state is unique up to overall 
\Uone transformations. Below $\Tz$ the ground state is degenerate 
and the time-reversal symmetry is broken.
}
\label{Fig:BTRS}
\end{figure}

It is well known that going through a phase transition allows 
uncorrelated regions to fall into different ground states 
\cite{Kibble:76,Zurek:85}. This is the Kibble-Zurek (KZ) 
mechanism for the formation of topological defects (see 
\cite{Rivers:01} for a review, for discussion 
in the context of chiral $p$-wave superconductor, see 
\cite{Vadimov.Silaev:13}). 
As different regions fall into either of the \Ztwo states,
domain walls are created while a superconductor goes through 
the transition to the broken \groupUZ state. \Figref{Fig:BTRS} 
shows the time-reversal symmetry breaking process while cooling 
down to $s+is$ state (for recent microscopic calculations 
of the appearance of $s+is$ state, see \cite{Maiti.Chubukov:13}).
Since their energy increases linearly with their length, closed 
domain walls contract and collapse or can be absorbed by 
boundaries. Here we propose a mechanism to stabilize domain 
walls, using geometrical barriers. 
We use numerical simulations that \emph{mimic} the KZ mechanism, 
to depict experimental set-ups to nucleate, stabilize and observe 
domain walls in $s+is$ state. 

In this work, we use the minimal Ginzburg-Landau (GL) free 
energy functional modeling a frustrated 
three-band superconductor
\Align{FreeEnergy}{
 \F=& \frac{\B^2}{2}
+\sum_{a=1}^3\frac{1}{2}|(\Grad+ie\A)\psia|^2 
+\alpha_a|\psia|^2+\frac{1}{2}\beta_a|\psia|^4 \nonumber \\
&-\sum_{a=1}^3\sum_{b>a}^3\etaab|\psia||\psib|\cos(\varphi_b-\varphi_a)\,.
}
The complex fields $\psia=|\psia| e^{i\varphi_a}$ in \Eqref{FreeEnergy} 
represent the superconducting condensates (labeled with $a,b$). 
They are electromagnetically coupled by the vector potential $\A$. 
And the coupling constant $e$ is used to parametrize the London 
penetration length of the magnetic field $\B=\bs\nabla\times\A$.
We model temperature dependence of the coefficients as
$\alpha_a\simeq\alphaZ_a\left( T/T_a -1\right)$ ($\alphaZ_a$ and 
$T_a$ being characteristic constants). We investigate only a 
limited range of temperature $T/\Tc\in[0.8;1]$, where $T_c$ is the 
common critical temperature. In general the GL coefficients have more 
complicated temperature dependencies (see \eg \cite{Silaev.Babaev:12}). 
However these dependencies are not very important for the questions 
which studied here.
Moreover, our results qualitatively should also apply beyond the GL 
regime. This is because, as shown in \cite{Maiti.Chubukov:13}, 
the GL model captures the overall structure of normal modes and 
length scales of the full microscopic theory of the $s+is$ state.
Thus, as long as the overall structure of the microscopically 
calculated phase diagram \cite{Maiti.Chubukov:13,Stanev.Tesanovic:10} 
is preserved, spontaneous breaking of the \groupUZ symmetry as well 
as domain wall formation should occur.

In the frustrated regime, when all three Josephson terms cannot 
simultaneously attain their optimal values and the resulting ground 
state phase differences $\varphiab\equiv\varphi_b-\varphi_a$ are 
neither $0$ nor $\pi$ \cite{Stanev.Tesanovic:10,Carlstrom.Garaud.ea:11a}. 
The ground state thus spontaneously breaks the time-reversal symmetry. 
For general consideration of phase locking between arbitrary number 
of components, see \cite{Weston.Babaev:13}.

As mentioned above, we model formation of domain walls during a 
cooling though \Ztwo phase transition. We explore different temperature 
dependent routes to the TRS breaking, predicted by microscopic theory 
\cite{Maiti.Chubukov:13,Stanev.Tesanovic:10}. The first route, which 
we refer to as set I (see \cite{Supplementary} for details and the 
chosen values of GL parameters), is the transition from the $s_{++}$ 
state to the $s+is$ state. There, the system goes from a three-band 
TRS state to the three-band BTRS. 
The alternative possibility, which we refer as set II, is the transition 
from the $s_\pm$ state to the $s+is$ state. That is, from a two-band (TRS) 
state to the three-band BTRS \cite{Maiti.Chubukov:13,Stanev.Tesanovic:10}. 
Since there are two discrete ground states, different regions of a 
frustrated superconductor with BTRS can fall in either the \Ztwo states 
and these regions are then separated by a domain wall. As a result, 
during BTRS phase transition (at $T=\Tz$), domain walls are created.
We consider field configurations varying in the $xy$ plane,  
with a normal magnetic field and assume translational invariance 
along $z$-direction.
A superconductor subject to an external field ${\bs H}=H\ez$
is described by the Gibbs free energy $\G=\F-\B\cdot {\bf H}$.
To evaluate the different responses 
the Gibbs free energy is minimized 
\footnote{ 
Note that KZ mechanism involves actual
time dependence. In our approach, we use a minimization algorithm 
instead of solving the actual time-dependent equations. At each 
temperature, once the algorithm has converged, the system is 
stationary. Then temperature is changed by certain amount 
$\delta T$ and minimization is repeated. Thus we do not simulate 
the actual Kibble-Zurek dynamical problem. Rather it is a 
quasi-equilibrium process which mimics the features of KZ mechanism. 
Our quasi-equilibrium simulation account for a number of features 
what would happen in the actual time-dependent evolution (such as 
spontaneous domain wall formation when the step $\delta T$ is 
sufficiently large, which corresponds to a rapid cooling). 
While we cannot predict rate for formation of topological defect
this simulation is sufficient to study the problem of geometric
 stabilization.
}
within a finite element framework provided by the {\tt Freefem++} 
library \cite{Hecht.Pironneau.ea}  
(for details, see the discussion in the 
Supplementary material \cite{
[{See Supplementary Material in appendix, for details of the 
parameters and numerical methods. Animations of the magnetization 
processes and field cooled experiments are also available at } ]
[{}]
Supplementary}).



\begin{figure}[!htb]
\hbox to \linewidth{ \hss
\includegraphics[width=.300\linewidth]{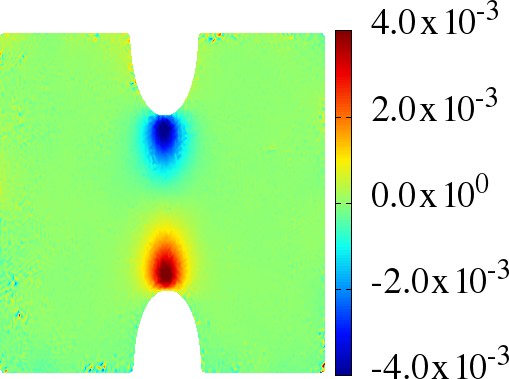}
\hspace{-0.15cm}\includegraphics[width=.265\linewidth]{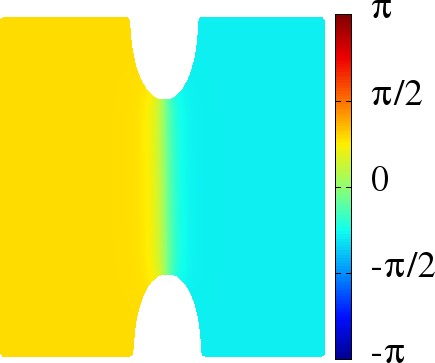}
\hspace{0.22cm} \includegraphics[width=.300\linewidth]{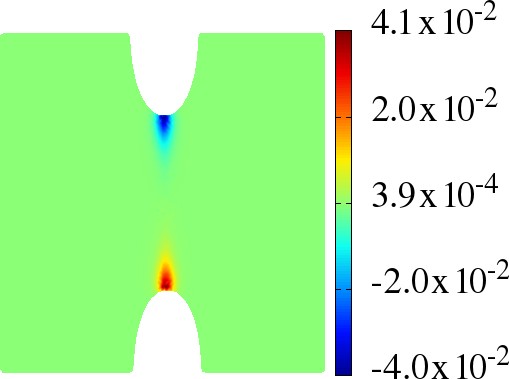}
\hss}
\vspace{0.25cm}
\hbox to \linewidth{ \hss
\includegraphics[width=.300\linewidth]{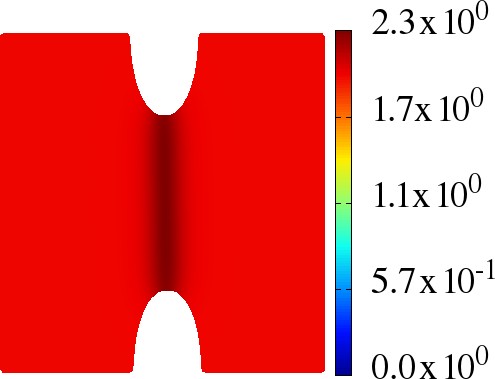}
\includegraphics[width=.300\linewidth]{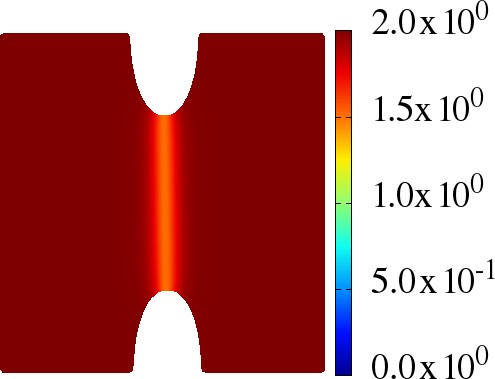}
\includegraphics[width=.300\linewidth]{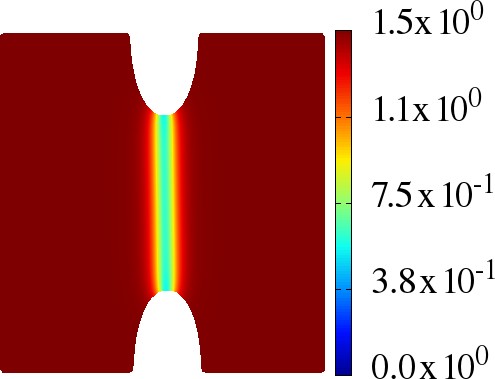}
\hss}
\setlength{\unitlength}{\linewidth}
\begin{picture}(1,0)
\put(0.145,0.5295){\makebox(0,0){\strut{}\footnotesize $\B$}}
\put(0.465,0.5295){\makebox(0,0){\strut{}\footnotesize $\varphi_{12}$}}
\put(0.760,0.5295){\makebox(0,0){\strut{}\footnotesize Counterflows}}
\put(0.150,0.27){\makebox(0,0){\strut{}\footnotesize $|\psi_1|^2$}}
\put(0.465,0.27){\makebox(0,0){\strut{}\footnotesize $|\psi_2|^2$}}
\put(0.770,0.27){\makebox(0,0){\strut{}\footnotesize $|\psi_3|^2$}}
\end{picture}
\caption{
(Color online) -- 
A geometrically stabilized domain wall in a non-convex domain, 
at $T/\Tc=0.8$ for the parameter set I. The domain wall is 
geometrically trapped, since to escape it should increase its 
length, which is energetically costly.
The phase difference $\varphi_{12}$ show that during the cooling, 
domain walls were created and one has been stabilized by the 
sample's geometry. The unfavourable phase differences 
at the domain wall affects the densities of the condensates. 
$|\psi_1|^2$ overshoots at the domain wall, while $|\psi_2|^2$ 
and $|\psi_3|^2$ are depleted. 
Note that the domain wall has a magnetic signature: spots of the 
dipole-like magnetic field, where the domain wall touches the bumps.  
It originates in features of the interband counterflow at the domain 
wall, discussed in the text. The upper right panel shows the contribution
to magnetic field of the second term in \Eqref{MagneticField}.
}
\label{Fig:Geometric-Stabilization}
\end{figure}
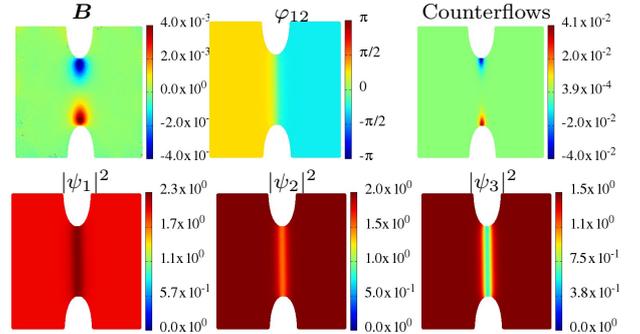 
While a frustrated superconductor is quenched through  $\Tz$, 
the temperature of the BTRS phase transition, domain walls 
are created. Because of their line tension, domain walls are 
unstable to be absorbed by the boundaries, or collapse if they 
are closed. 
Here we propose a mechanism for stabilization of domain walls, 
by using a geometric barrier. Such a barrier exists if a sample 
has a non-convex geometry as for example shown on 
\Figref{Fig:Geometric-Stabilization}. 
Next we will show, that when a domain wall is stabilized it 
has experimentally detectable features that can signal $s+is$ 
state. As shown in \Figref{Fig:Geometric-Stabilization}, 
if during a quench a domain wall ending on non-convex bumps is 
created, it can relax to a stable configuration.
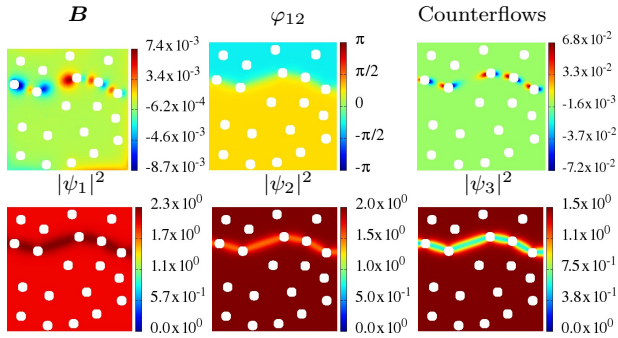
\begin{figure}[!htb]
\hbox to \linewidth{ \hss
\includegraphics[width=.300\linewidth]{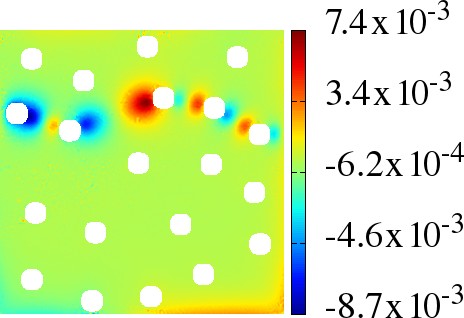}
\hspace{-0.05cm}\includegraphics[width=.265\linewidth]{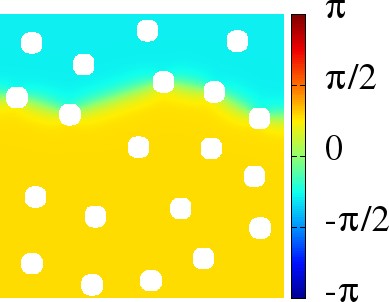}
\hspace{0.24cm} \includegraphics[width=.300\linewidth]{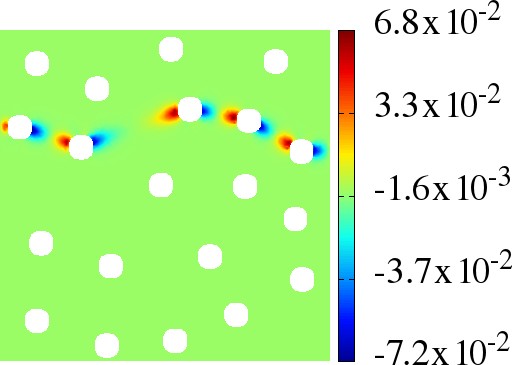}
\hss}
\vspace{0.27cm}
\hbox to \linewidth{ \hss
\includegraphics[width=.300\linewidth]{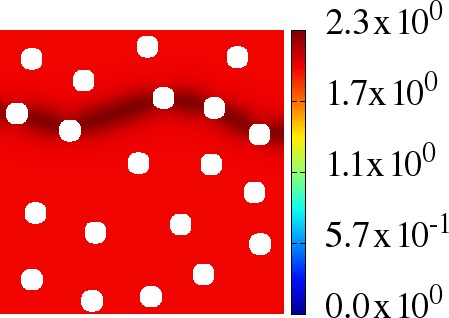}
\includegraphics[width=.300\linewidth]{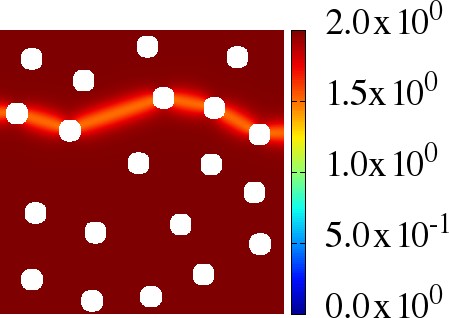}
\includegraphics[width=.300\linewidth]{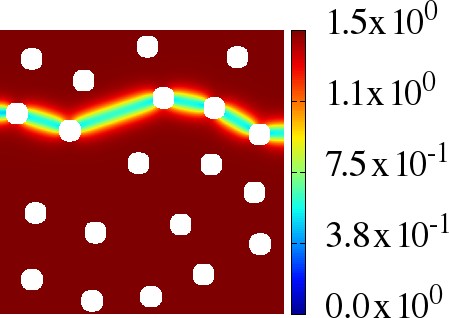}
\hss}
\setlength{\unitlength}{\linewidth}
\begin{picture}(1,0)
\put(0.145,0.5295){\makebox(0,0){\strut{}\footnotesize $\B$}}
\put(0.465,0.5295){\makebox(0,0){\strut{}\footnotesize $\varphi_{12}$}}
\put(0.760,0.5295){\makebox(0,0){\strut{}\footnotesize Counterflows}}
\put(0.150,0.27){\makebox(0,0){\strut{}\footnotesize $|\psi_1|^2$}}
\put(0.465,0.27){\makebox(0,0){\strut{}\footnotesize $|\psi_2|^2$}}
\put(0.770,0.27){\makebox(0,0){\strut{}\footnotesize $|\psi_3|^2$}}
\end{picture}
\caption{
(Color online) -- 
A domain wall stabilized by randomly located pinning centers. 
When cooled past $\Tz$, domain walls are formed at random positions. 
Then, during the relaxation process, the quench-induced domain wall 
is stabilized against collapse by nearby pinning centers.
Displayed quantities and physical parameters are the same as in
\Figref{Fig:Geometric-Stabilization}. 
Here again, the domain wall has a dipole-like signature of the 
magnetic field where it is attached to the pinning 
centers.  
}
\label{Fig:Pinning-Stabilization}
\end{figure} 
Indeed, to join its ends and collapse to zero size, the domain 
wall would have to increase its length first, it is thus in a 
stable equilibrium while trapped on the bumps. 
Exactly the same effect is present when there is a pinning by 
inhomogeneities instead of a geometric barrier (see 
\Figref{Fig:Pinning-Stabilization}). This kind of pinning 
induces similar magnetic dipole signatures.

To simulate the cooling experiment, the energy is minimized 
at $T=\Tc+\dT$, \ie starting in the normal state. The temperature 
is subsequently decreased with a step $\dT$ and the energy minimized 
for the new temperature (\ie new $\alpha_a$'s).
The faster the system undergoes a phase transition, the more defects 
are nucleated. This is achieved, in our simulations, by cooling with 
bigger temperature steps (see animations in \cite{Supplementary} for 
a typical domain wall-stabilizing process). Domain walls are always 
created, but their location is random and thus they do not always 
geometrically stabilize.
We performed several simulations of the cooling processes and 
verified that indeed the number of produced defects is larger 
when temperature steps are bigger. Conversely, to ensure that 
no DW is formed, the system has to be cooled very slowly.

Remarkably, as shown in Figs.~\ref{Fig:Geometric-Stabilization} and 
\ref{Fig:Pinning-Stabilization}, even 
in zero applied field the domain wall carries opposite, non-zero 
magnetic field at its ends. Yet the total net flux through the 
sample is zero. The magnitude of this effect depends on the width 
of the domain wall (and thus on the parameters of the model).
For \Figref{Fig:Geometric-Stabilization}, the amplitude of the 
local fields is of the order of magnitude of a percent of the 
magnetic field of a vortex. 
The origin of this signature in $s+is$ state is principally 
different from magnetic signature of domain walls in $p+ip$ 
superconductor. 
Namely, in $p+ip$ superconductors, DW carry uniform magnetic 
field originating from orbital momentum of Cooper pairs 
(see e.g. \cite{goldbart,raghu, Vadimov.Silaev:13}). 
Here, by contrast, the domain walls carry magnetic field 
only where they are attached to the boundary and the field
inverts its direction so that there is no net flux. 
This magnetic field originates in interband counterflow 
in the presence of relative density gradients. Indeed, 
the magnetic field has the following dependence on the 
field gradients \cite{Garaud.Carlstrom.ea:13}:
\Align{MagneticField}{
B_z&=-\epsilon_{ij}\partial_i\left(\frac{J_j}{e^|\Psi|^2}\right) \nonumber \\
&-\frac{i\epsilon_{ij}}{e^2|\Psi|^4} \left[
   |\Psi|^2\partial_i\Psi^\dagger\partial_j\Psi
   +\Psi^\dagger\partial_i\Psi\partial_j\Psi^\dagger\Psi
   \right] \,,
}
with $\Psi^\dagger=(\psi_1^*,\psi_2^*,\psi_3^*)$ and 
$|\Psi|^2=\Psi^\dagger\Psi$. The interband counterflow contribution 
to $\B$ is the second term \Eqref{MagneticField}. That is, density 
gradients mixed with gradients of phase differences (see 
Figs.~\ref{Fig:Geometric-Stabilization} and 
\ref{Fig:Pinning-Stabilization}). In the total magnetic field 
signature, counterflows are partially screened by the first term.

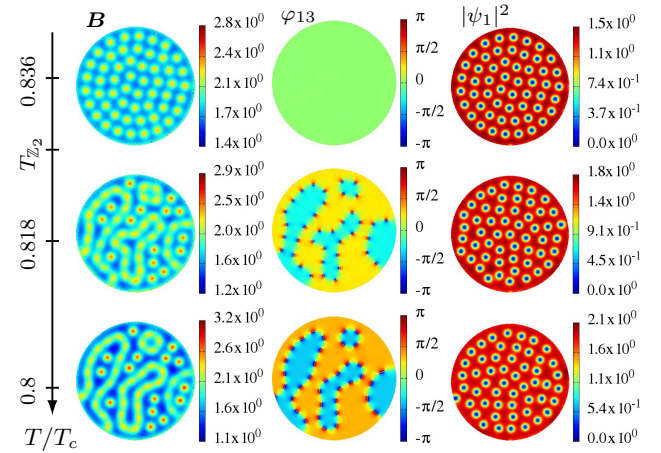
\begin{figure}[!htb]
\minipage{.05\linewidth}
\setlength{\unitlength}{\linewidth}
\begin{picture}(1,12)
\linethickness{0.25mm}
\put(0.5,10.5){\rotatebox{-270}{\makebox(0,0){\strut{}\footnotesize $0.836$}}}
\put(1,10.6){\line(1,0){0.4}}
\put(0.5,8.3){\rotatebox{-270}{\makebox(0,0){\strut{}\footnotesize $\Tz$}}}
\put(1,8.4){\line(1,0){0.4}}
\put(0.5,5.5){\rotatebox{-270}{\makebox(0,0){\strut{}\footnotesize $0.818$}}}
\put(1,5.6){\line(1,0){0.4}}
\put(0.5,1){\rotatebox{-270}{\makebox(0,0){\strut{}\footnotesize $0.8$}}}
\put(1,1.1){\line(1,0){0.4}}
\put(1.2,12){\vector(0,-1){11.75}}
\put(1.05,-0.5){\makebox(0,0){\strut{} $T/\Tc$}}
\end{picture}
\endminipage
\minipage{.95\linewidth}
	\hbox to \linewidth{ \hss
	\includegraphics[width=.300\linewidth]{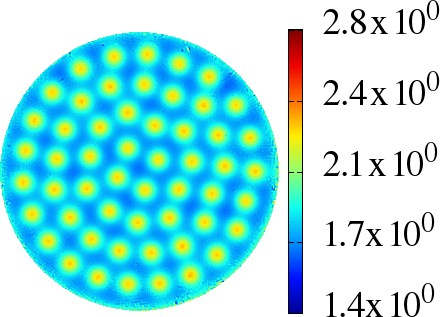}
	\includegraphics[width=.275\linewidth]{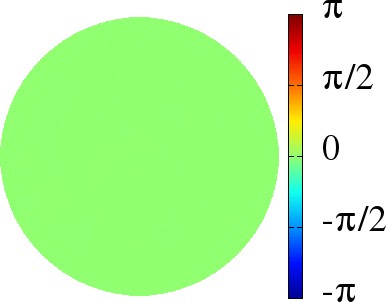}
	\includegraphics[width=.300\linewidth]{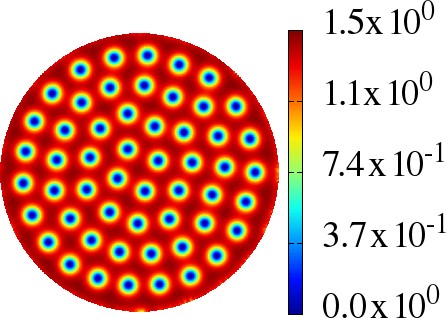}
	\hss}
\vspace{0.15cm}
	\hbox to \linewidth{ \hss
	\includegraphics[width=.300\linewidth]{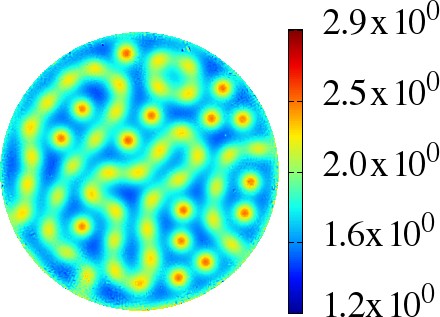}
	\includegraphics[width=.275\linewidth]{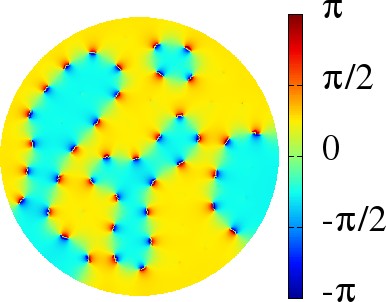}
	\includegraphics[width=.300\linewidth]{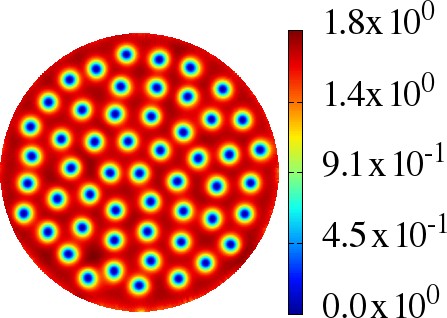}
	\hss}
\vspace{0.15cm}
	\hbox to \linewidth{ \hss
	\includegraphics[width=.300\linewidth]{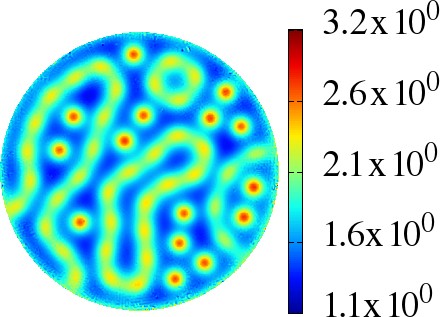}
	\includegraphics[width=.275\linewidth]{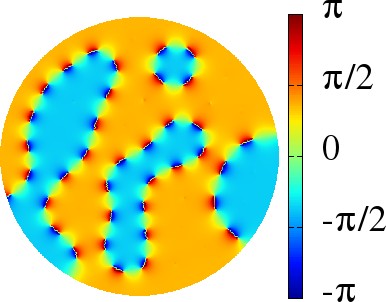}
	\includegraphics[width=.300\linewidth]{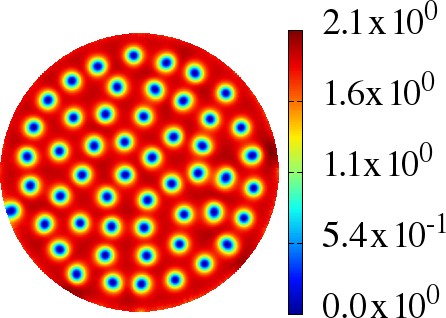}
	\hss}
\endminipage
\begin{picture}(1,0)
\put(-215,80){\makebox(0,0){\strut{}\footnotesize $\B$}}
\put(-139,83){\makebox(0,0){\strut{}\footnotesize $\varphi_{13}$}}
\put(-69,82){\makebox(0,0){\strut{}\footnotesize $|\psi_1|^2$}}
\end{picture}
\caption{
(Color online) -- 
Field cooled experiment for the parameter set II, under an applied 
field $H/\Phi_0S=70$. 
First the system is a two-band ($\psi_2=0$) and thus it is TRS. 
Then when cooled through $\Tz$ ($\psi_2\neq0$) the system enters 
the BTRS regime and different region pick up different ground state 
phase-locking. The resulting DW are stabilized against complete 
contraction by the already existing vortices.
}
\label{Fig:Field-cooled}
\end{figure}

For modeling a field cooled experiment, the Gibbs energy for a 
given applied field $H$ is minimized for decreasing temperatures. 
This is shown in \Figref{Fig:Field-cooled}. 
At $\Tc$ superconductivity sets in and the sample is filled with 
vortices. Then while temperature is further decreased, past \Ztwo 
phase transition (at $\Tz$), KZ mechanism leads to the formation 
of domain walls. 
As shown in \Figref{Fig:Field-cooled}, the pre-existing 
vortices stabilize the domain wall against collapse (regardless 
of the geometry). These domain walls either terminate on the 
boundary or are closed.
Closed domain walls stabilized by vortices were considered in 
\cite{Garaud.Carlstrom.ea:11,Garaud.Carlstrom.ea:13}.
These are Skyrmions since they are characterized by \CPtwo 
topological invariant. 
Note that to accommodate the unfavourable phase differences at 
the DW, it is beneficial to split vortices into three types of 
fractional vortices (see detailed discussion in \cite{Garaud.Carlstrom.ea:11,Garaud.Carlstrom.ea:13}). 
Since at the DW, there is less total density, the penetration 
length is effectively smaller and vortices appear bigger. The 
DW can clearly be identified when measuring the magnetic field.

Consider now  magnetization process at fixed $T<\Tz$. No field 
is initially applied ($H=0$) and the superconductor is in one 
ground state. The applied field is increased with a step $\dH$. 
There are no preexisting DW and, as long as the applied field is 
below \HcO, no vortex enters  the system. The Meissner state
survives to fields higher than \HcO  because of the Bean-Livingston 
barrier. While the applied field is further increased, vortices 
enter and arrange in a triangular lattice.
Note that, big steps $\dH$ can provide enough energy to locally 
fall into the opposite \Ztwo state during a relaxation process. 
This thus leads to the formation of a domain wall which is 
stabilized by the presence of vortices (see \cite{Supplementary}).

\begin{figure}[!htb]
	\hbox to \linewidth{ \hss
\minipage{.05\linewidth}
\setlength{\unitlength}{\linewidth}
\begin{picture}(1,21)
\linethickness{0.25mm}
\put(0.5,20.1){\rotatebox{-270}{\makebox(0,0){\strut{}\footnotesize $18$}}}
\put(1,20.1){\line(1,0){0.4}}
\put(0.5,17.0){\rotatebox{-270}{\makebox(0,0){\strut{}\footnotesize \HcO}}}
\put(1,17.1){\line(1,0){0.4}}
\put(0.5,15.4){\rotatebox{-270}{\makebox(0,0){\strut{}\footnotesize $75$}}}
\put(1,15.4){\line(1,0){0.4}}
\put(0.5,10.6){\rotatebox{-270}{\makebox(0,0){\strut{}\footnotesize $87$}}}
\put(1,10.6){\line(1,0){0.4}}
\put(0.5,5.75){\rotatebox{-270}{\makebox(0,0){\strut{}\footnotesize $108$}}}
\put(1,5.75){\line(1,0){0.4}}
\put(0.5,1.1){\rotatebox{-270}{\makebox(0,0){\strut{}\footnotesize $135$}}}
\put(1,1.1){\line(1,0){0.4}}
\put(1.2,21.75){\vector(0,-1){21.750}}
\put(1.05,-1.0){\makebox(0,0){\strut{} $\frac{H}{\Phi_0S}$}}
\end{picture}
\endminipage
\minipage{.95\linewidth}
	\hbox to \linewidth{ \hss
	\includegraphics[width=.300\linewidth]{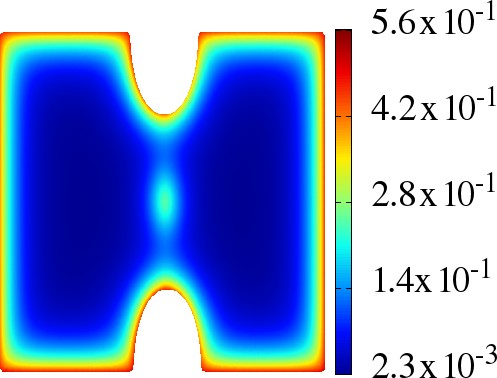}
	\includegraphics[width=.275\linewidth]{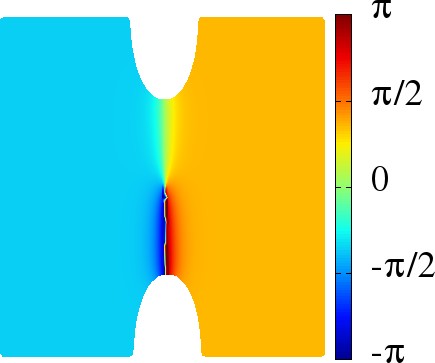}
	\includegraphics[width=.300\linewidth]{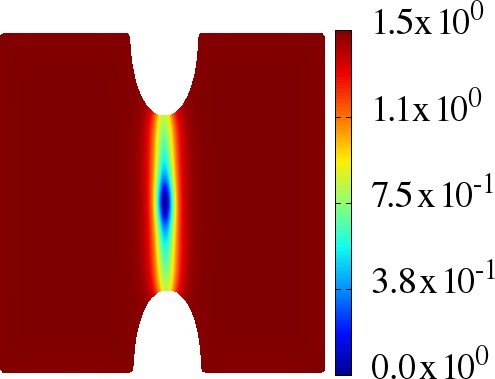}
	\hss}
\vspace{0.15cm}
	\hbox to \linewidth{ \hss
	\includegraphics[width=.300\linewidth]{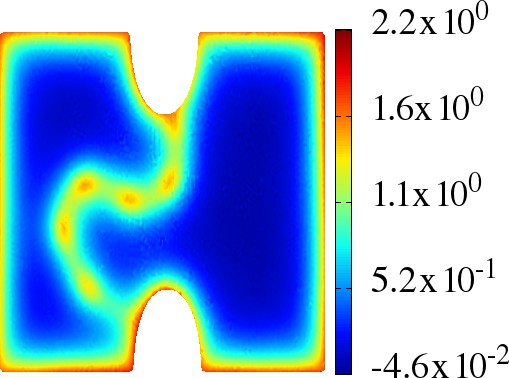}
	\includegraphics[width=.275\linewidth]{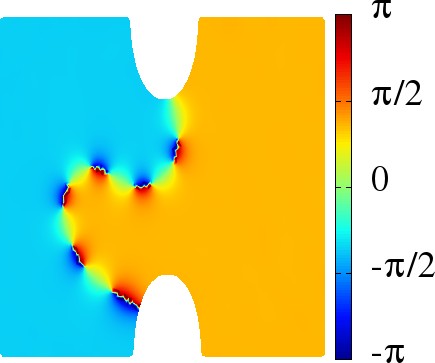}
	\includegraphics[width=.300\linewidth]{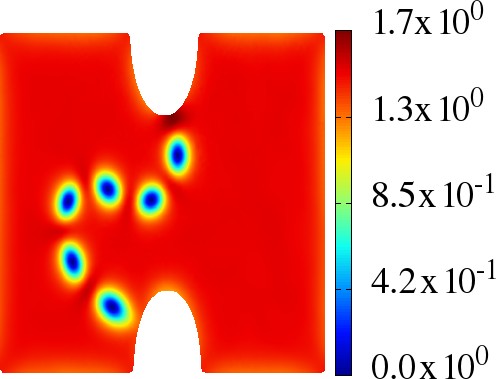}
	\hss}
\vspace{0.15cm}
	\hbox to \linewidth{ \hss
	\includegraphics[width=.300\linewidth]{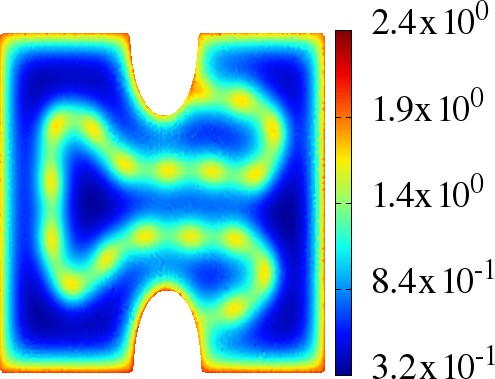}
	\includegraphics[width=.275\linewidth]{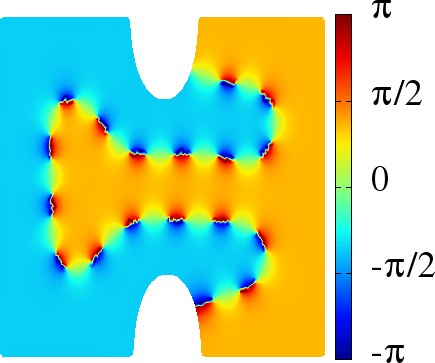}
	\includegraphics[width=.300\linewidth]{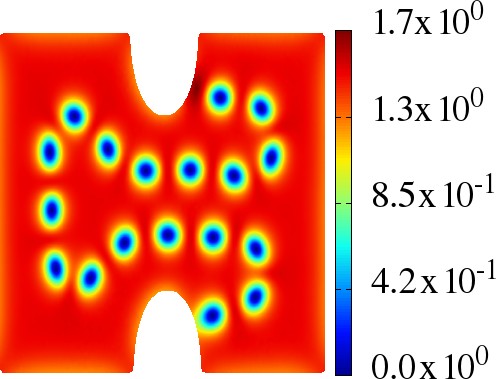}
	\hss}
\vspace{0.15cm}
	\hbox to \linewidth{ \hss
	\includegraphics[width=.300\linewidth]{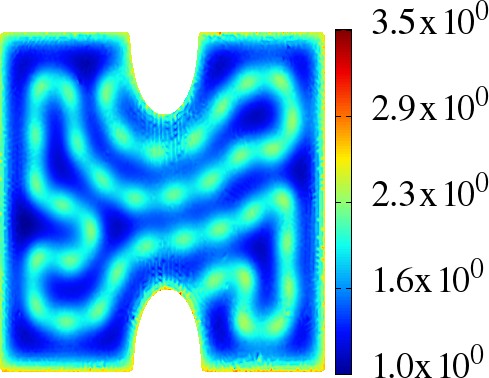}
	\includegraphics[width=.275\linewidth]{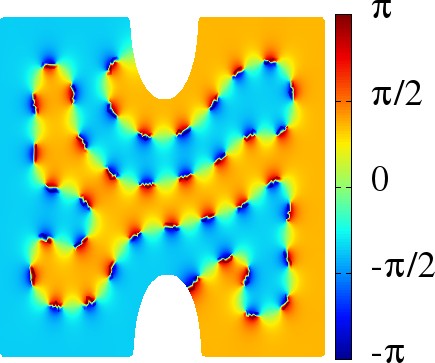}
	\includegraphics[width=.300\linewidth]{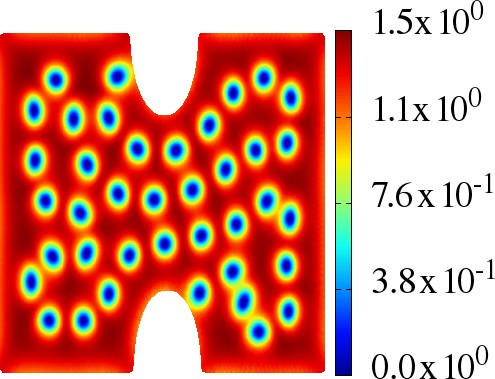}
	\hss}
\vspace{0.15cm}
	\hbox to \linewidth{ \hss
	\includegraphics[width=.300\linewidth]{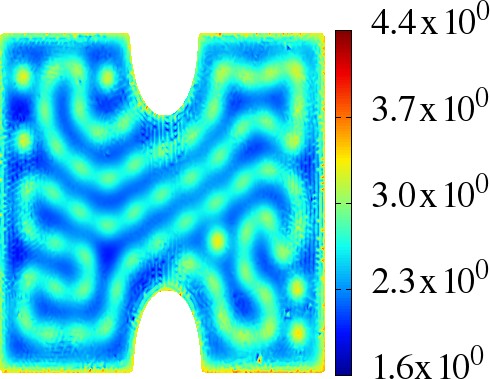}
	\includegraphics[width=.275\linewidth]{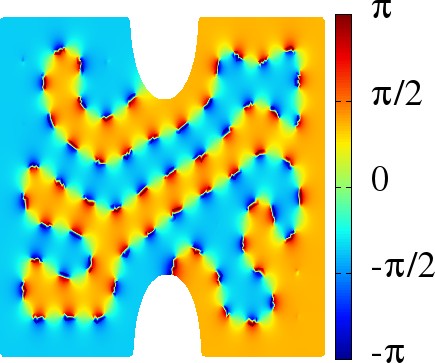}
	\includegraphics[width=.300\linewidth]{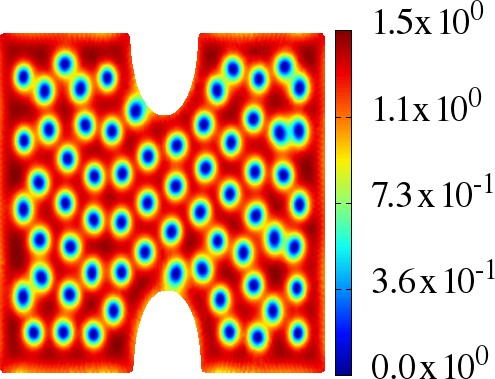}
	\hss}
\endminipage
\hss}
\begin{picture}(1,0)
\put(-74,290){\makebox(0,0){\strut{}\footnotesize $\B$}}
\put(0,293){\makebox(0,0){\strut{}\footnotesize $\varphi_{13}$}}
\put(70,292){\makebox(0,0){\strut{}\footnotesize $|\psi_3|^2$}}
\end{picture}
\hbox to \linewidth{ \hss
\resizebox{\linewidth}{!}{\input{figure-5P}}
\hss}
\vspace{-0.8cm}
\caption{
(Color online) -- 
Magnetization process of a three-band BTRS, when the zero field 
configuration has the geometrically stabilized domain wall
\Figref{Fig:Geometric-Stabilization} (all parameters being the same). 
First vortex entry, way below \HcO, shown in the top row is a 
fractional vortex in $\psi_3$. This can be seen from the phase 
difference $\varphi_{13}$ which winds $2\pi$. 
The red curve is the magnetization curve of this process, 
while the blue curve is a reference magnetization, starting 
from a uniform ground state. 
}
\label{Fig:Magnetization-DW}
\end{figure}

Now we consider the regime of our main interest.
As shown in \Figref{Fig:Magnetization-DW}, the magnetization process 
in the presence of quench-induced and geometrically stabilized domain 
wall is very unusual. Since some density components are depleted at 
the domain wall (see \Figref{Fig:Geometric-Stabilization}), vortex 
entry for the corresponding component costs much less energy there 
than from the boundaries. The first vortex entry occurs at much lower 
fields than \HcO. Here, a core is created   only in one band, thus it is 
a {\it fractional} vortex which enters the domain wall. Fractional 
vortices are thermodynamically unstable in a uniform bulk superconducting 
state because they have logarithmically divergent energy 
\cite{Garaud.Carlstrom.ea:13}. The situation here is different 
because the sample has a pre-existing domain wall. See 
\cite{Supplementary} for all quantities. 
In increased field the domain wall is filled with vortices. Despite 
its energy cost, it eventually becomes beneficial to elongate the 
domain wall. It starts bending and gradually fills the sample. 
At the first integer vortex entry, the sample is already filled 
with the flux-carrying DW.
The associated magnetization curves also show striking differences 
from the case without domains-wall. 
This can provide a way to confirm $s+is$ superconductivity. For a 
sample whose geometry allows stabilization of DW, magnetization 
process after a rapid cooling (or other kind of thermal quench) 
can be significantly different from that of the same, slowly cooled 
sample. The first will show magnetization process different from 
the reference measurement. Chances to stabilize domain walls are 
further enhanced by having multiple stabilizing geometric barriers.

In conclusion, we have studied domain walls in $s+is $ 
superconductors. We presented proposal for an experimental 
set-up which can lead to formation of stable domain walls. 
We demonstrated that domain walls in $s+is$ superconductors 
have magnetic signatures which could be  detected in 
scanning SQUID, Hall, or magnetic force microscopy measurements.
Moreover we showed that for geometrically stabilized DW, 
the magnetization curve could change substantially as DW 
allows flux penetration in the form of fractional vortices 
in low fields. Thus a sample subject to different cooling 
processes should exhibit very different magnetization process 
and magnetization curves.

The observation of these features can signal  $s+is$ state 
(because in contrast $s_{\pm}$ and $s_{++}$ states do not break 
\Ztwo symmetry and thus have no domain walls), for example in 
hole-doped Ba$_{1-x}$K$_x$Fe$_2$As$_2$ \cite{Maiti.Chubukov:13}.

\begin{acknowledgments}
We acknowledge fruitful discussions with J.~Carlstr\"om and 
D.~Weston. 
This work is supported by the Swedish Research Council, by the 
Knut and Alice Wallenberg Foundation through the Royal Swedish 
Academy of Sciences fellowship and by NSF CAREER Award No. 
DMR-0955902.
The computations were performed on resources provided by the 
Swedish National Infrastructure for Computing (SNIC) at National 
Supercomputer Center at Linkoping, Sweden.
\end{acknowledgments}


%

\clearpage
\appendix
\setcounter{section}{1}
\setcounter{equation}{0}
\renewcommand{\theequation}{\Alph{section}.\arabic{equation}}

\section{Phase diagram in the Ginzburg-Landau regime}

\begin{figure}[!htb]
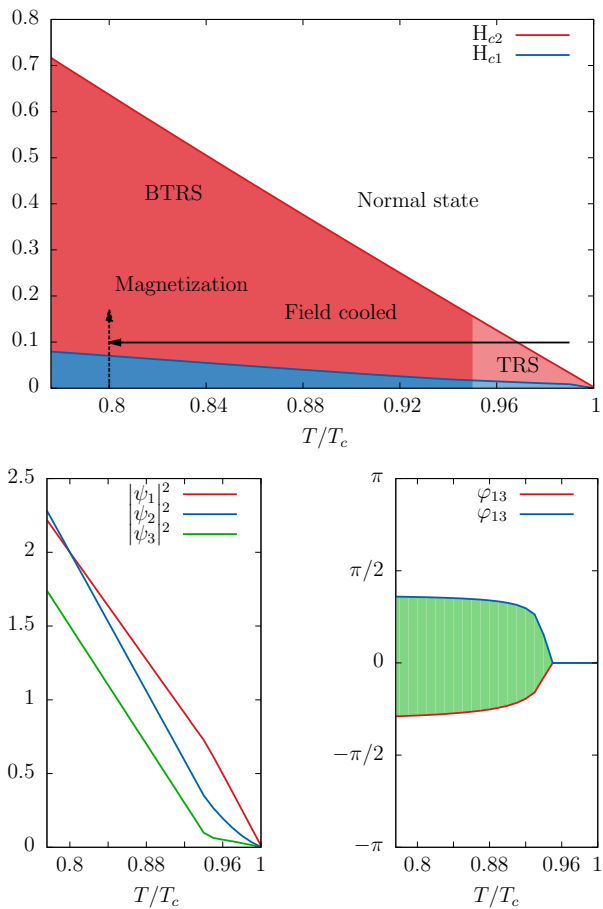

\hbox to \linewidth{ \hss
\resizebox{\linewidth}{!}{\input{figure-supp-1A}}
\hss}
\hbox to \linewidth{ \hss
\resizebox{.5\linewidth}{!}{\input{figure-supp-1B}}
\resizebox{.5\linewidth}{!}{\input{figure-supp-1C}}
\hss}
\caption{
(Color online) -- 
Phase diagram and ground state configuration for the parameter 
set I. That is, the Ginzburg-Landau parameters are 
$(\alphaZ_1,\alphaZ_2,\alphaZ_3)=(8.273,10.412,8.5)$ and 
$(T_1,T_2,T_3)=(0.91,0.885,0.85)$. The 
gauge coupling $e=0.3$, while $\beta_a=1$ and 
$\eta_{12}=\eta_{13}=-\eta_{23}=2$. 
}
\label{Fig:AppPhaseDiag1}
\end{figure}

\begin{figure}[!htb]
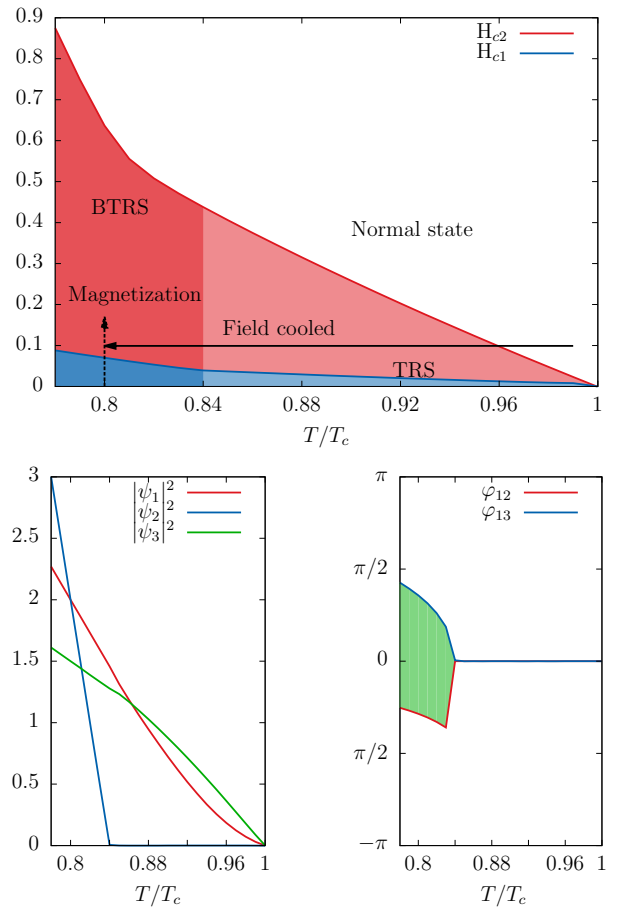

\hbox to \linewidth{ \hss
\resizebox{\linewidth}{!}{\input{figure-supp-2A}}
\hss}
\hbox to \linewidth{ \hss
\resizebox{.5\linewidth}{!}{\input{figure-supp-2B}}
\resizebox{.5\linewidth}{!}{\input{figure-supp-2C}}
\hss}
\caption{
(Color online) -- 
Phase diagram and ground state configuration for the parameter set II. 
That is, the Ginzburg-Landau parameters are 
$(\alphaZ_1,\alphaZ_2,\alphaZ_3)=(11.8108,41,4.9444)$ and 
$(T_1,T_2,T_3)=(0.874,0.82,0.89)$. The gauge coupling $e=0.3$, while 
$\beta_a=1$ and $\eta_{12}=\eta_{13}=-\eta_{23}=2$. 
}
\label{Fig:AppPhaseDiag2}
\end{figure}

For our purposes, we do not need to reproduce phase diagram of
\cite{Maiti.Chubukov:13} quantitatively. It is sufficient, 
to retain temperature dependency only of the following coefficients:
\Equation{AppAlpha}{
\alpha_a\equiv\alpha_a(T)
=\alphaZ_a\ln\frac{T}{T_a}
\simeq\alphaZ_a\left( \frac{T}{T_a} -1\right) \,.
}
The temperatures are scaled so that in zero applied field, $\Tc=1$. 
We investigate only a restricted range of temperatures $T/\Tc\in[0.8;1]$.
Figures \ref{Fig:AppPhaseDiag1} and \ref{Fig:AppPhaseDiag2} display 
the $H/T$ diagram for the parameter sets studied in the main text. 
The actual values of the parameters are given in the captions.
In the main text, we discuss two possible routes to break time-reversal 
symmetry during the cooling process. Such a phase diagram agrees 
with microscopic calculations \cite{Maiti.Chubukov:13} and is 
reproduced here phenomenologically in the framework of a minimal 
Ginzburg-Landau model.

For the first symmetry breaking 
pattern, which we refer as set I, superconductivity arises for all 
three-bands below $\Tc$. The frustration becomes important enough to 
break time-reversal symmetry only below $\Tz<\Tc$. Thus before reaching 
the BTRS state, the system is a three-band TRS superconductor 
(the $s_{++}$ state). The related phase diagram is displayed in 
\Figref{Fig:AppPhaseDiag1}.
The second symmetry breaking pattern which we investigate, shown in 
\Figref{Fig:AppPhaseDiag2} is different. For temperatures 
$\Tz<T<\Tc$, only two bands develop superconductivity (the $s_\pm$ state). 
There thus is only one Josephson term, and the phases are trivially locked. 
Below $\Tz$, the third condensate becomes superconducting and the 
time-reversal symmetry is broken (the $s+is$ state) because all three 
(frustrated) Josephson terms are important enough.

Note that at $T=0.8$, the values of the parameters of the Ginzburg-Landau 
functional are the same for parameter sets I and II. The magnetization 
process at $T=0.8$ is thus the same for both these systems.

\section{Finite element energy minimization} \label{AppNumerics}

We consider the two-dimensional problem is defined 
on the bounded domain $\Omega\subset\mathbbm{R}^2$ with $\partial\Omega$ 
its boundary. $\Omega$ can assume any geometry. In particular, it can 
be convex (homeomorph to a disc) or non-convex. The problem is 
supplemented by the boundary condition ${\bs n}\cdot\D\psi_a=0$ with 
${\bs n}$ the normal vector to $\partial\Omega$. Physically this 
condition implies there is no current flowing through the boundary.
A superconductor subject to an external field ${\bs H}=H\ez$ is described 
by the Gibbs free energy $\G=\F-\B\cdot {\bs H}$, yielding the boundary 
conditions on $\partial\Omega$ for the vector potential $\Curl\A={\bs H}$. 

The variational problem is defined for numerical computation using a 
finite element formulation provided by the {\tt Freefem++} library 
\cite{Hecht.Pironneau.ea}. Discretization within finite element 
formulation is done via a (homogeneous) triangulation over $\Omega$, 
based on Delaunay-Voronoi algorithm. Functions are decomposed on a 
continuous piecewise quadratic basis over each triangle. 
The accuracy of such method is controlled through the number of 
triangles, (we typically used $3\sim6\times10^4$), the order of 
expansion of the basis on each triangle (2nd order polynomial basis 
on each triangle), and also the order of the quadrature formula for 
the integral on the triangles. 

Once the problem is posed, a numerical optimization algorithm is 
used to solve the variational non-linear problem (\ie to find the 
minima of $\G$). We used here a non-linear conjugate gradient method.
The algorithm is iterated until relative variation of the norm of 
the gradient of the functional  $\G$ with respect to all degrees of 
freedom is less than $10^{-6}$.

\subsection{Field cooled experiments}

Simulating a field cooled experiment, is done through the following 
sequences. For a given value of the applied field $H$, the initial 
temperature is chosen so that it slightly exceeds the second critical 
field ($H>\Hc{2}$). Then the Gibbs energy is minimized for a given 
temperature. For the next step  the temperature is decreased by
step $\dT$ and the Gibbs energy is subsequently minimized using 
solution at the previous temperature as an initial guess.
To ensure that the system is not trapped into an artificial minimum, 
a small white noise corresponding to small thermal fluctuations is 
added at each temperature step, before further relaxing the energy. 
This procedure, corresponds to an horizontal path in the $H(T)$ 
diagram. It is iterated down to a given temperature, which we 
chose to be $T_{min}=0.8\Tc$.

\begin{figure}[!htb]
\hbox to \linewidth{ \hss
\minipage{.05\linewidth}
\setlength{\unitlength}{\linewidth}
\begin{picture}(1,12)
\linethickness{0.25mm}
\put(0.5,10.6){\rotatebox{-270}{\makebox(0,0){\strut{}\footnotesize $129.5$}}}
\put(1,10.6){\line(1,0){0.4}}
\put(0.5,5.75){\rotatebox{-270}{\makebox(0,0){\strut{}\footnotesize $140$}}}
\put(1,5.75){\line(1,0){0.4}}
\put(0.5,1.1){\rotatebox{-270}{\makebox(0,0){\strut{}\footnotesize $143.5$}}}
\put(1,1.1){\line(1,0){0.4}}
\put(1.2,11.5){\vector(0,-1){11.15}}
\put(1.05,-0.5){\makebox(0,0){\strut{} $\frac{H}{\Phi_0S}$}}
\end{picture}
\endminipage
\minipage{.95\linewidth}
	\hbox to \linewidth{ \hss
	\includegraphics[width=.300\linewidth]{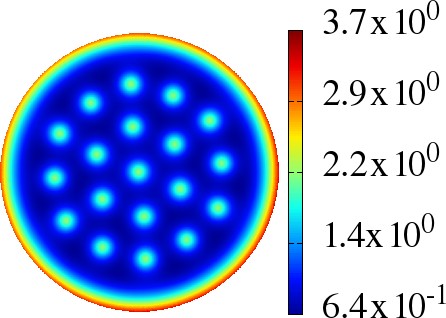}
	\includegraphics[width=.275\linewidth]{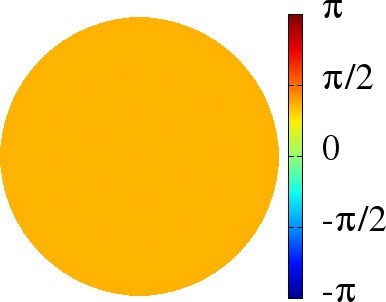}
	\includegraphics[width=.300\linewidth]{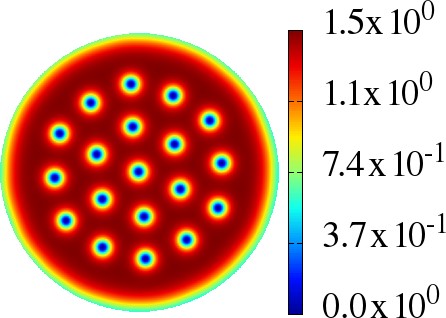}
	\hss}
\vspace{0.15cm}
	\hbox to \linewidth{ \hss
	\includegraphics[width=.300\linewidth]{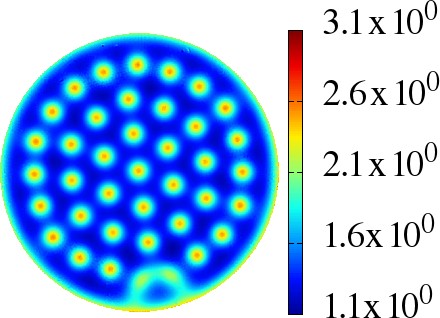}
	\includegraphics[width=.275\linewidth]{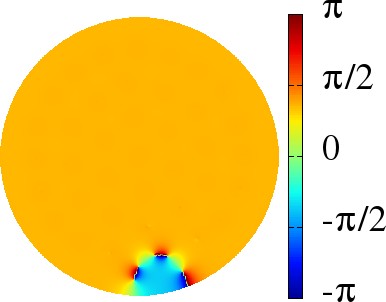}
	\includegraphics[width=.300\linewidth]{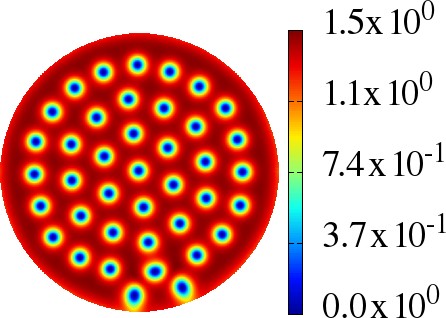}
	\hss}
\vspace{0.15cm}
	\hbox to \linewidth{ \hss
	\includegraphics[width=.300\linewidth]{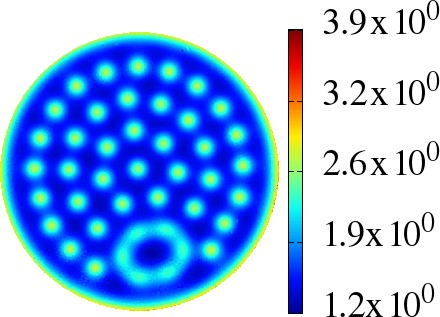}
	\includegraphics[width=.275\linewidth]{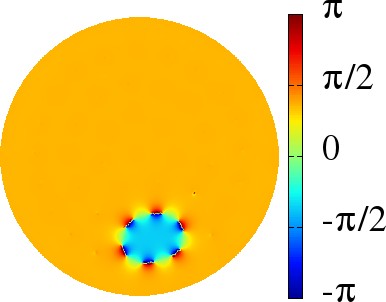}
	\includegraphics[width=.300\linewidth]{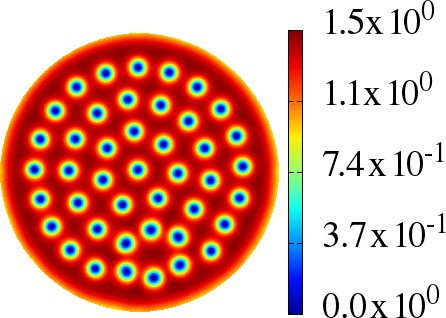}
	\hss}
\endminipage
\hss}
\begin{picture}(1,0)
\put(-89,160){\makebox(0,0){\strut{}\footnotesize $\B$}}
\put(-15,163){\makebox(0,0){\strut{}\footnotesize $\varphi_{13}$}}
\put(55,162){\makebox(0,0){\strut{}\footnotesize $|\psi_3|^2$}}
\end{picture}
\caption{
(Color online) -- 
Magnetization process of a three-band superconductor in the BTRS 
state. As the domain is convex, the zero field configuration has no 
domain wall.
Increase of the applied field can lead the system to locally fall 
into the opposite \Ztwo state and seeds a domain wall. The latter 
is stabilized by the vortices as in Fig.~3 of the paper. 
The parameter are the same as in Fig.~2 of the main text. 
}
\label{Fig:AppMagnetization-noDW}
\end{figure}
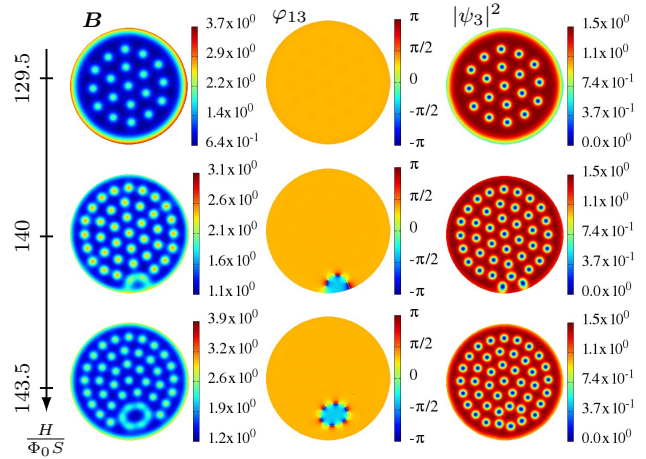

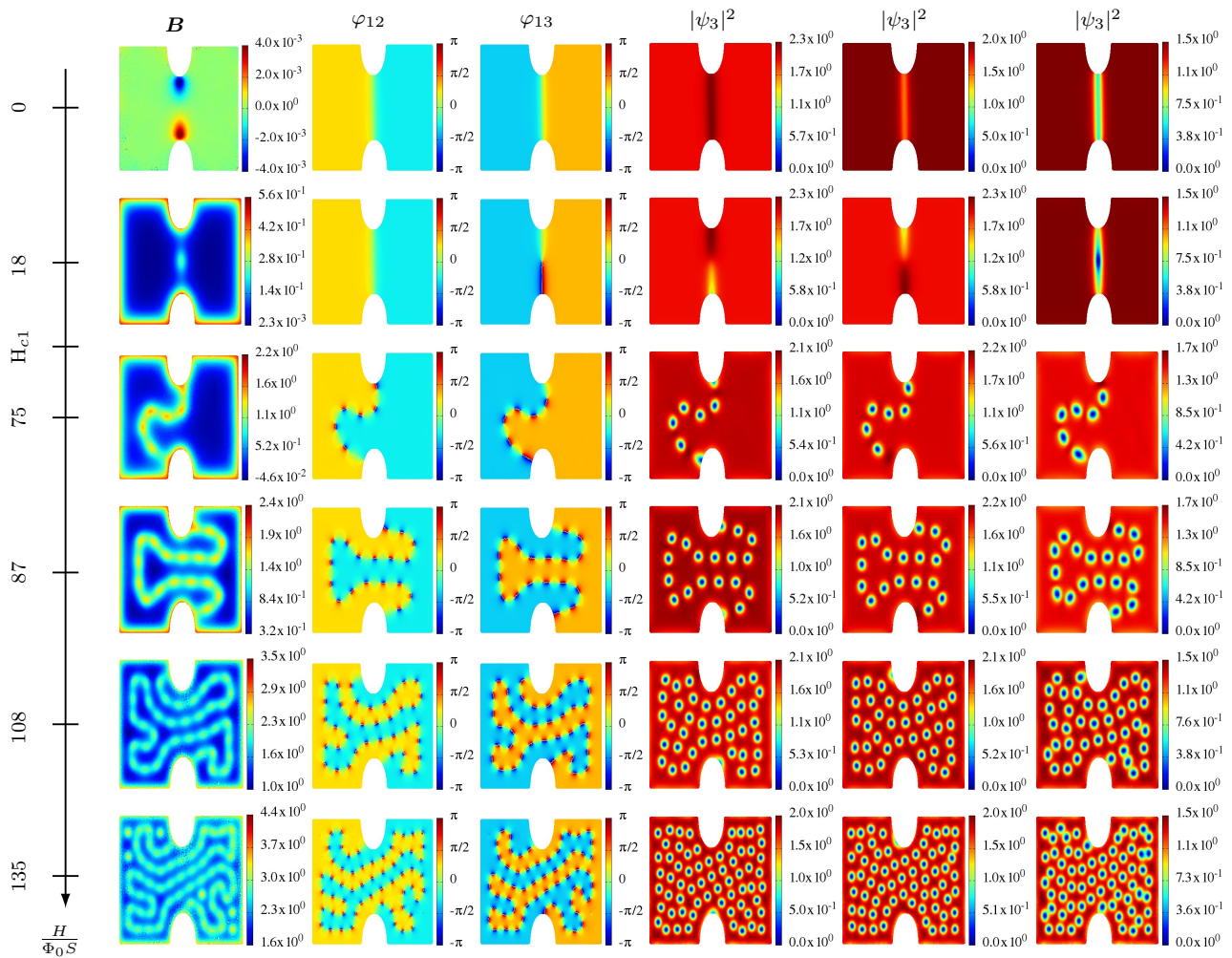
\begin{figure*}[!htb]
\hbox to \linewidth{ \hss
\minipage{.05\linewidth}
\setlength{\unitlength}{\linewidth}
\begin{picture}(1,15)
\linethickness{0.25mm}
\put(0.5,13.40){\rotatebox{-270}{\makebox(0,0){\strut{}\footnotesize $0$}}}
\put(1,13.40){\line(1,0){0.4}}
\put(0.5,10.95){\rotatebox{-270}{\makebox(0,0){\strut{}\footnotesize $18$}}}
\put(1,11.0){\line(1,0){0.4}}
\put(0.5,9.7){\rotatebox{-270}{\makebox(0,0){\strut{}\footnotesize \HcO}}}
\put(1,9.7){\line(1,0){0.4}}
\put(0.5,8.6){\rotatebox{-270}{\makebox(0,0){\strut{}\footnotesize $75$}}}
\put(1,8.6){\line(1,0){0.4}}
\put(0.5,6.2){\rotatebox{-270}{\makebox(0,0){\strut{}\footnotesize $87$}}}
\put(1,6.2){\line(1,0){0.4}}
\put(0.5,3.80){\rotatebox{-270}{\makebox(0,0){\strut{}\footnotesize $108$}}}
\put(1,3.850){\line(1,0){0.4}}
\put(0.5,1.5){\rotatebox{-270}{\makebox(0,0){\strut{}\footnotesize $135$}}}
\put(1,1.5){\line(1,0){0.4}}
\put(1.2,14.0){\vector(0,-1){13.0}}
\put(1.05,0.50){\makebox(0,0){\strut{} $\frac{H}{\Phi_0S}$}}
\end{picture}
\endminipage
\minipage{.95\linewidth}
	\hbox to \linewidth{ \hss
	\includegraphics[width=.15\linewidth]{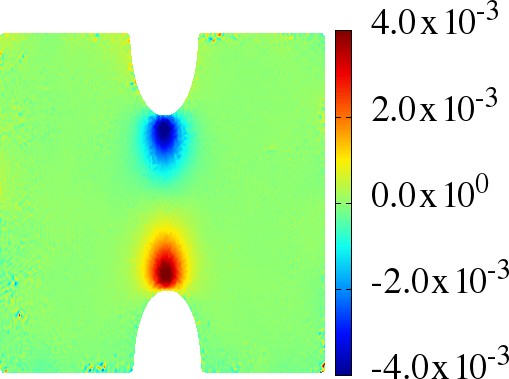}
	\includegraphics[width=.13\linewidth]{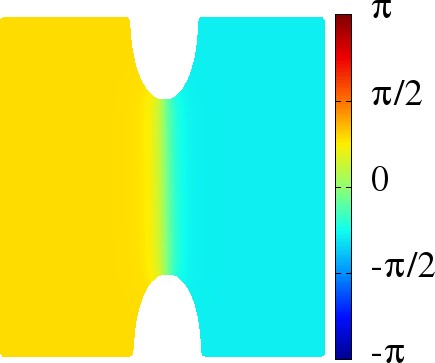}
	\includegraphics[width=.13\linewidth]{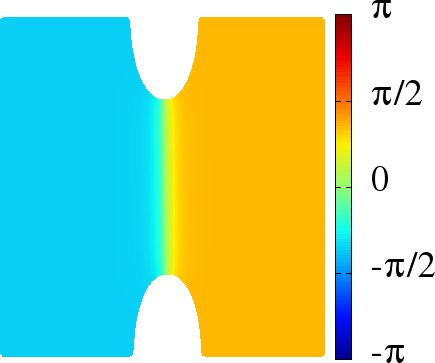}
	\includegraphics[width=.15\linewidth]{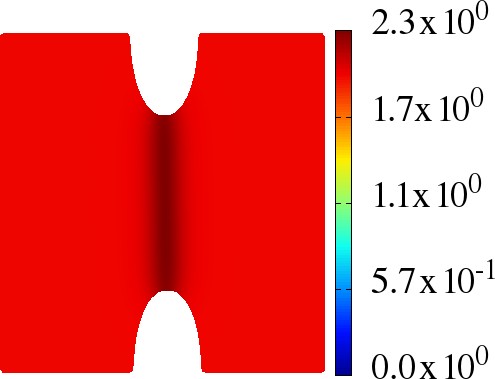}
	\includegraphics[width=.15\linewidth]{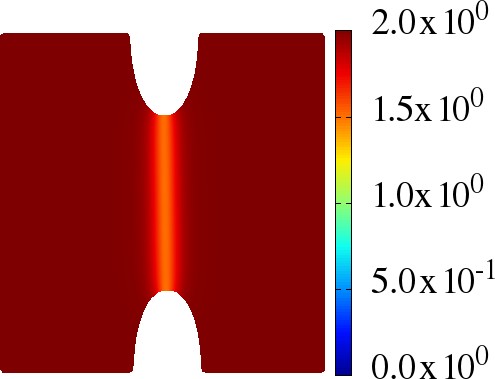}
	\includegraphics[width=.15\linewidth]{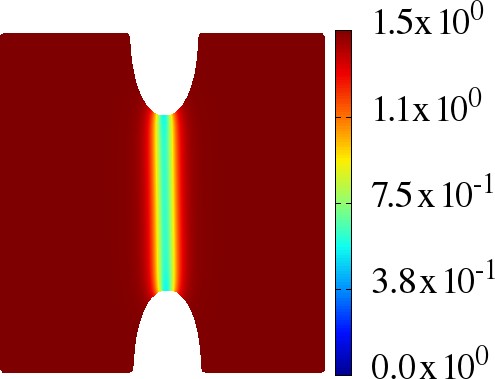}
	\hss}
\vspace{0.15cm}
	\hbox to \linewidth{ \hss
	\includegraphics[width=.15\linewidth]{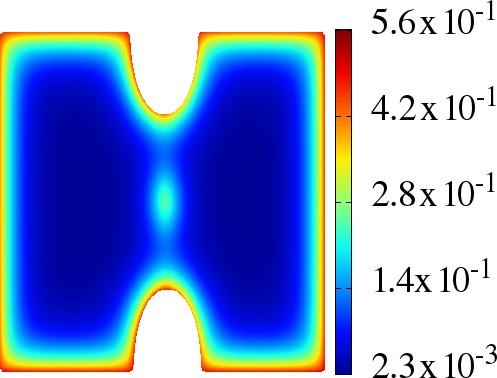}
	\includegraphics[width=.13\linewidth]{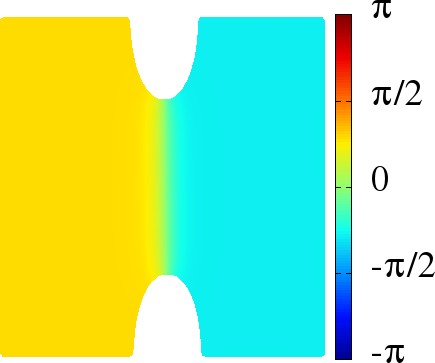}
	\includegraphics[width=.13\linewidth]{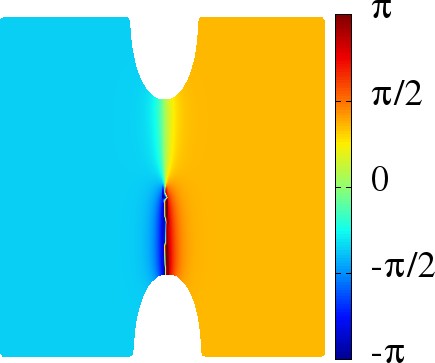}
	\includegraphics[width=.15\linewidth]{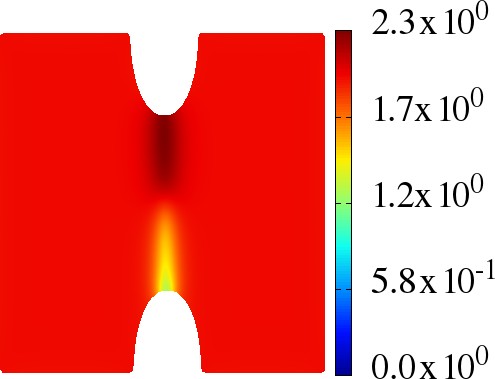}
	\includegraphics[width=.15\linewidth]{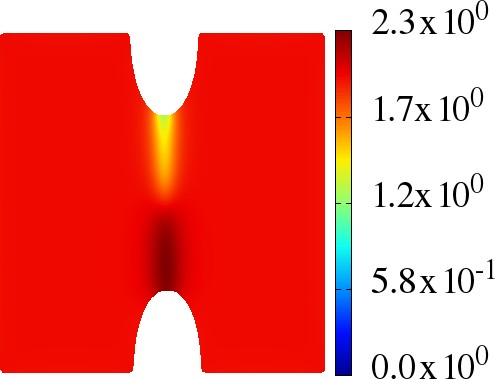}
	\includegraphics[width=.15\linewidth]{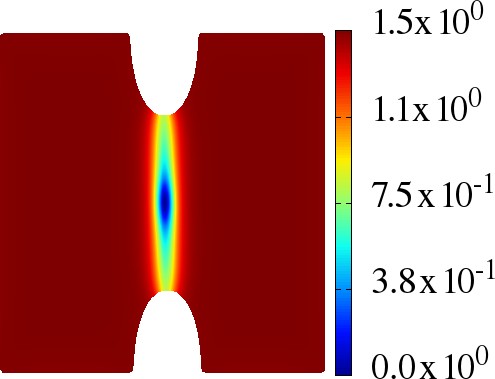}
	\hss}
\vspace{0.15cm}
	\hbox to \linewidth{ \hss
	\includegraphics[width=.15\linewidth]{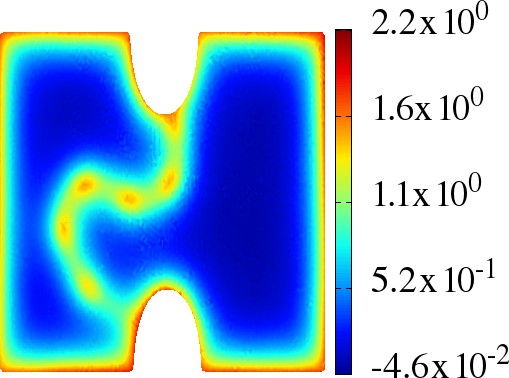}
	\includegraphics[width=.13\linewidth]{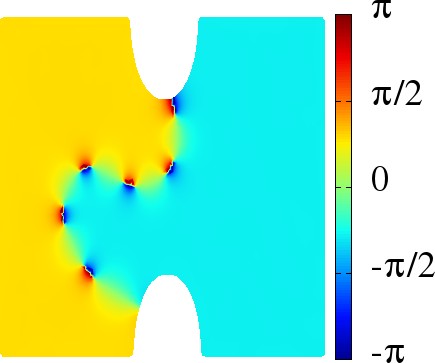}
	\includegraphics[width=.13\linewidth]{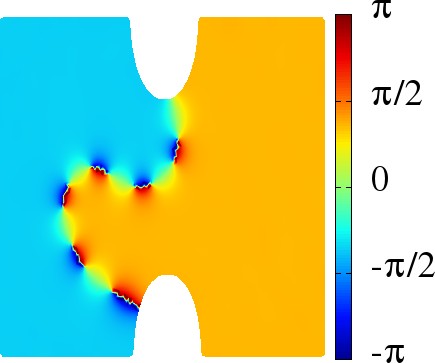}
	\includegraphics[width=.15\linewidth]{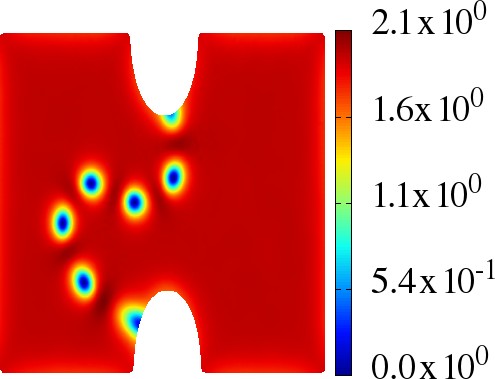}
	\includegraphics[width=.15\linewidth]{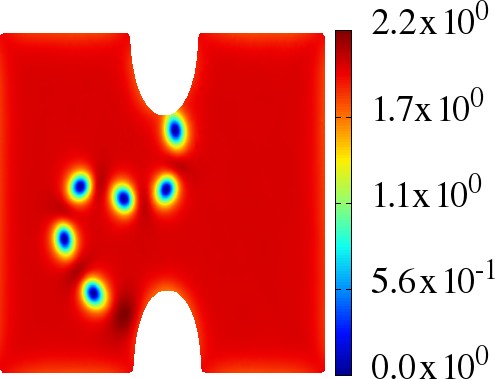}
	\includegraphics[width=.15\linewidth]{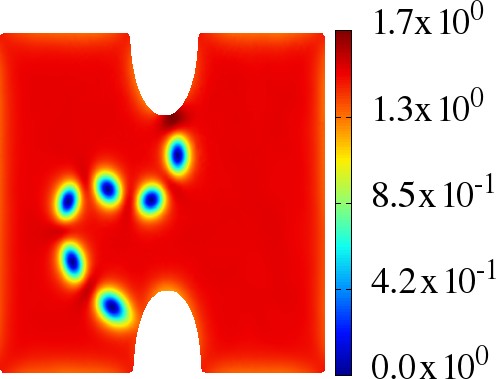}
	\hss}
\vspace{0.15cm}
	\hbox to \linewidth{ \hss
	\includegraphics[width=.15\linewidth]{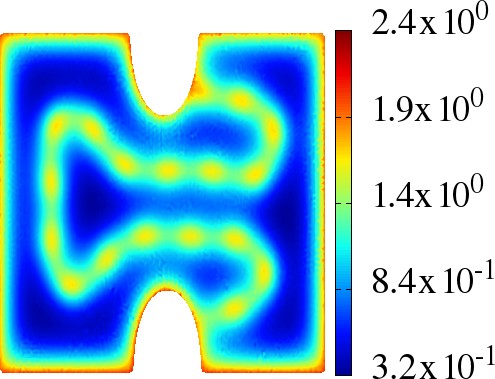}
	\includegraphics[width=.13\linewidth]{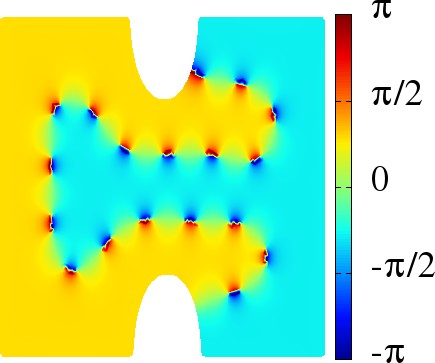}
	\includegraphics[width=.13\linewidth]{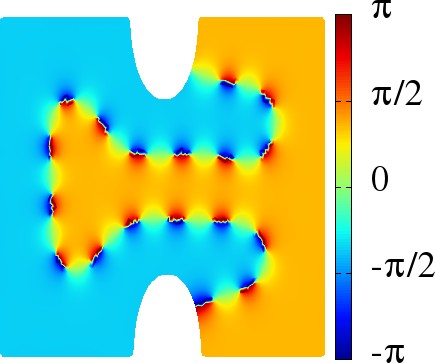}
	\includegraphics[width=.15\linewidth]{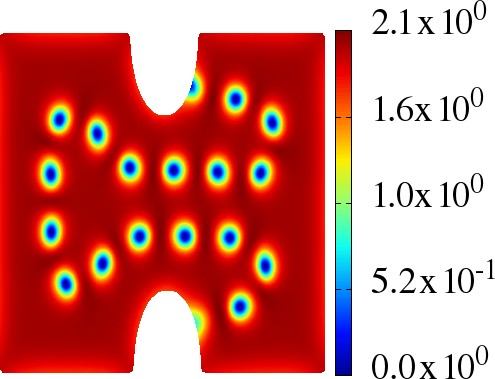}
	\includegraphics[width=.15\linewidth]{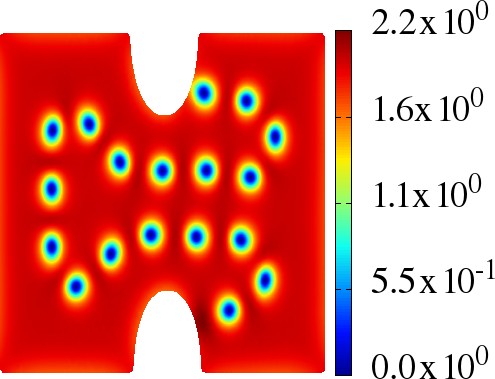}
	\includegraphics[width=.15\linewidth]{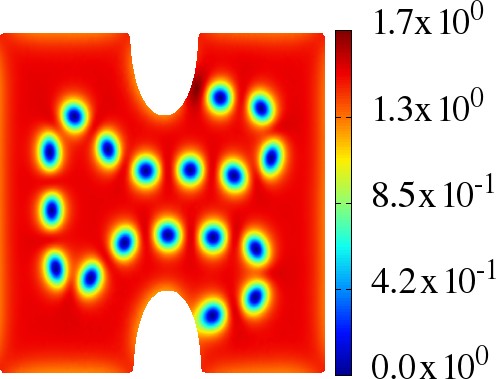}
	\hss}
\vspace{0.15cm}
	\hbox to \linewidth{ \hss
	\includegraphics[width=.15\linewidth]{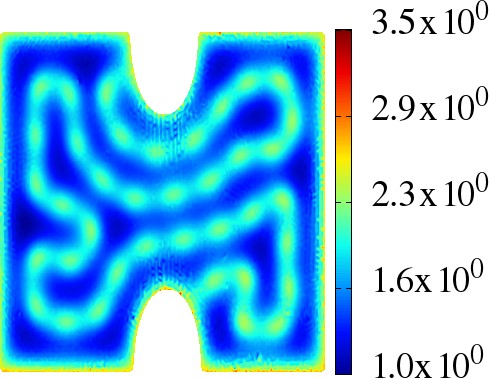}
	\includegraphics[width=.13\linewidth]{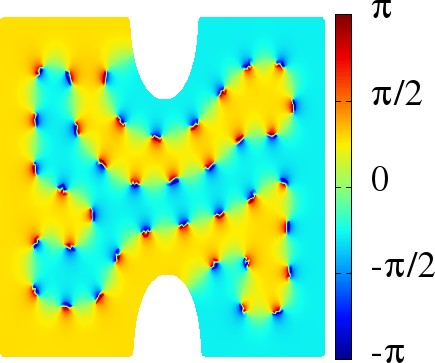}
	\includegraphics[width=.13\linewidth]{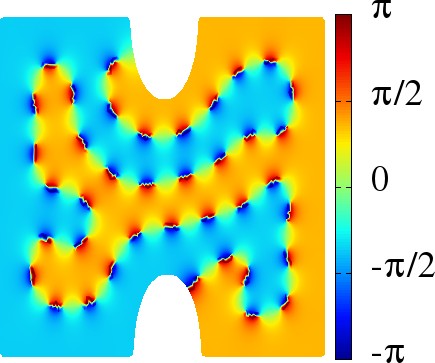}
	\includegraphics[width=.15\linewidth]{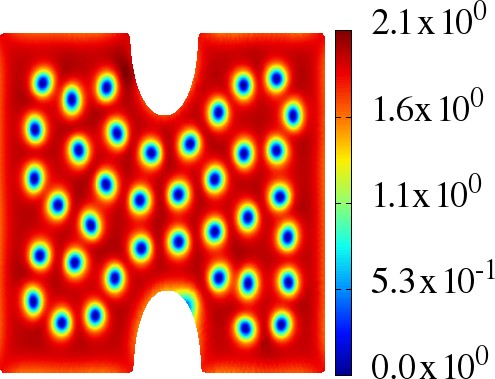}
	\includegraphics[width=.15\linewidth]{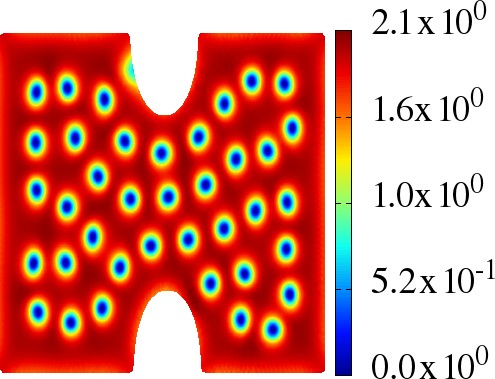}
	\includegraphics[width=.15\linewidth]{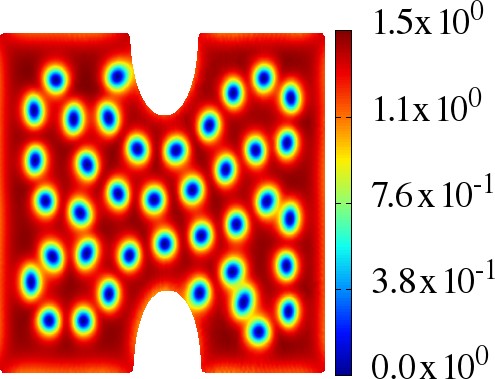}
	\hss}
\vspace{0.15cm}
	\hbox to \linewidth{ \hss
	\includegraphics[width=.15\linewidth]{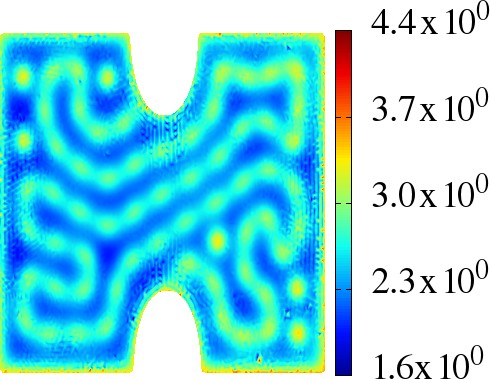}
	\includegraphics[width=.13\linewidth]{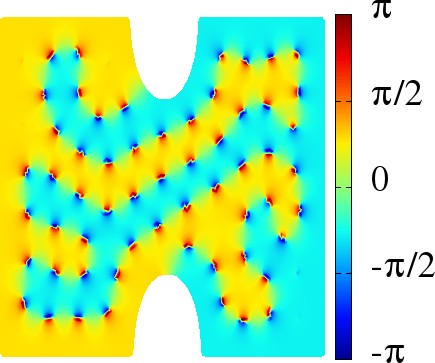}
	\includegraphics[width=.13\linewidth]{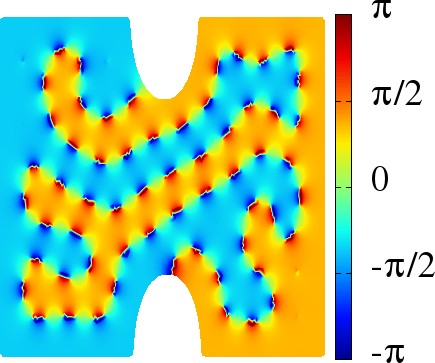}
	\includegraphics[width=.15\linewidth]{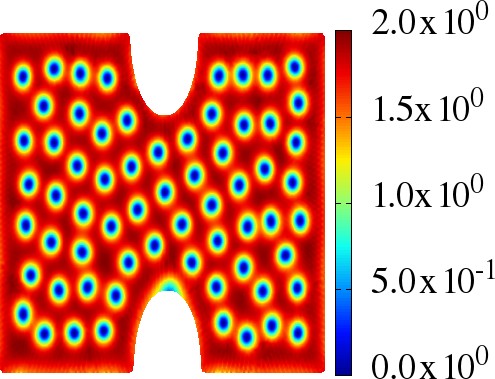}
	\includegraphics[width=.15\linewidth]{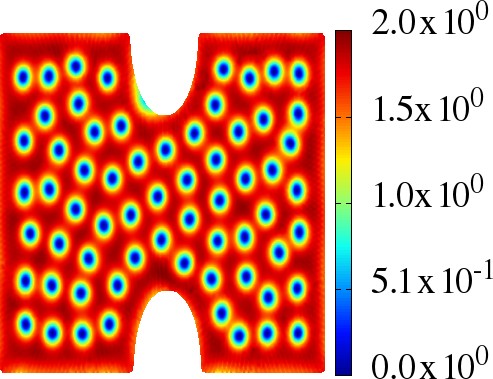}
	\includegraphics[width=.15\linewidth]{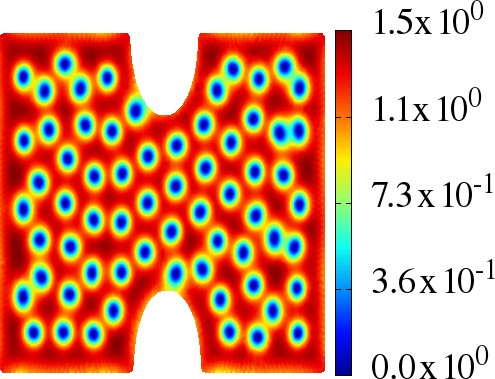}
	\hss}
\endminipage
\hss}
\begin{picture}(1,0)
\put(-178,375){\makebox(0,0){\strut{}\footnotesize $\B$}}
\put(-102,378){\makebox(0,0){\strut{}\footnotesize $\varphi_{12}$}}
\put(-36,378){\makebox(0,0){\strut{}\footnotesize $\varphi_{13}$}}
\put(33,377){\makebox(0,0){\strut{}\footnotesize $|\psi_3|^2$}}
\put(108,377){\makebox(0,0){\strut{}\footnotesize $|\psi_3|^2$}}
\put(184,377){\makebox(0,0){\strut{}\footnotesize $|\psi_3|^2$}}
\end{picture}
\vspace{-0.6cm}
\caption{
(Color online) -- 
Magnetization process of a three-band BTRS, when the zero field 
configuration has the geometrically stabilized domain wall
This displays additional quantities to the magnetization process
shown in Fig.~4 of the main text. 
First vortex entry, way below \HcO, shown in the second row is a 
fractional vortex in $\psi_3$. This can be seen from the phase 
difference $\varphi_{13}$ which winds $2\pi$ and that $|\psi_1|^2$ 
and $|\psi_2|^2$ have no singularity.
}
\label{Fig:AppMagnetization-DW}
\end{figure*}

Physically, this correspond to start the experiment with some applied 
field ${\bs H}$ at a temperature above the critical temperature.
That is, initially there is no superconducting state. Then while 
decreasing the temperature, normal state is no longer stable and the 
system goes to superconducting state with vortices (when there is 
a non-zero applied field). While cooled, the system goes across 
the BTRS transition at $\Tz$. There, different regions can fall into 
different ground states, thus leading to domain wall formation. This 
is Kibble-Zurek mechanism. Note that KZ mechanism involves actual
time dependence. In our approach, we use a minimization algorithm 
instead of solving the actual time-dependent equations. At each 
temperature, once the algorithm has converged, the system is 
stationary. Thus we do not simulate the actual dynamics of 
Kibble-Zurek. Rather it is a quasi-equilibrium process which mimics 
the KZ mechanism. 

\subsection{Magnetization process at fixed \texorpdfstring{$T$}{T}.}

This experiment investigates the response of the superconductor 
of an applied external field, at a fixed temperature. Below its 
critical temperature, no field is initially applied ($H=0$). 
The $H=0$ configuration is generated by cooling the sample from 
$\Tc$ to a preferred temperature below $\Tz$. Thus, during the 
cooling process domain walls have been created. For a convex 
geometry, they can always decay to zero, while they can be 
geometrically stabilized, for non-convex geometries.
For the magnetization process, the superconductor is initially 
either in the uniform ground state or non-uniform in the presence 
of a domain wall.

Then keeping the temperature fixed, the applied field is increased 
with a step $\dH$. The configuration of the condensates and vector 
potential of the previous step is used and Gibbs energy minimized 
for the new value of the applied field.
This corresponds physically to apply an increasing magnetic field 
at fixed temperature. As long as the applied field is below \HcO, 
it is not energetically preferable to nucleate flux carrying 
topological defects and the superconductor stays in the Meissner 
state. Above \HcO topological defects start to enter the system. 
Actually first vortex entry occurs for higher values of the applied 
field, since they have to overcome the Bean-Livingston barrier 
which depends on the geometry. 

Big steps $\dH$ in the applied field, can provide enough energy 
to locally fall into the opposite \Ztwo state during a relaxation 
process. As seen in \Figref{Fig:AppMagnetization-noDW}, this thus 
leads to the formation of a domain wall which is stabilized by the 
presence of vortices.

Additional quantities, to the unusual magnetization process in 
presence of a domain wall in zero applied field are displayed in 
\Figref{Fig:AppMagnetization-DW}.


\end{document}